\documentclass[aps,prl,twocolumn,superscriptaddress,address,amsmath,showpacs]{revtex4}

\usepackage{graphicx}

\begin{document}


\title{Strong surface contribution to the Nonlinear Meissner Effect}

\author{A. Zare, T. Dahm, and N. Schopohl}

\affiliation{Institut f\"ur Theoretische Physik and Center for 
Collective Quantum Phenomena, Universit\"at  T\"ubingen, Auf der Morgenstelle 14, D-72076 T\"ubingen, Germany}

\date{\today}

\begin{abstract}
We demonstrate that in a $d$-wave superconductor the bulk nonlinear Meissner effect
is dominated by a surface effect due to Andreev bound states at low temperatures.
The contribution of this surface effect to the nonlinear response coefficient follows
a $1/T^3$ law with opposite sign compared to the bulk $1/T$ behavior. The 
cross-over from bulk dominated behavior to surface dominated
behavior occurs at a temperature of $T/T_c \sim 1/\sqrt{\kappa}$.
We present an approximate analytical calculation, which supports our
numerical calculations and provides a qualitative understanding of the effect.
The effect can be probed by intermodulation distortion experiments.
\end{abstract}


\maketitle

In a superconductor with nodes in the gap function, quasi-particles
near the gap nodes lead to an intrinsic nonlinear electromagnetic
response \cite{Yip}. In a $d$-wave superconductor this nonlinear Meissner effect appears as
a linear magnetic field dependence of the magnetic penetration
depth at low temperatures \cite{Yip,Xu}, but can more
sensitively be probed by temperature dependent intermodulation distortion or
harmonic generation experiments \cite{Dahm}. The nonlinear
response coefficient shows an upturn at low temperatures
following a $1/T$ law in a clean system down to temperatures of
the order of $1/\kappa$, where $\kappa$ is the Ginzburg-Landau
parameter of the superconductor. This behavior has been confirmed
by intermodulation distortion experiments on high-$T_c$ cuprate superconductors
\cite{Jutzi,Oates,Booth}. At even lower temperatures nonlocal effects
\cite{Li}, as impurity effects \cite{Dahmimp}, lead to a
saturation of this low temperature upturn.

So far, theoretical studies of the nonlinear Meissner effect did
not consider the special electronic structure that appears at
the surface of a $d$-wave superconductor. At a surface that
has a finite angle with the (100) direction of the crystal,
Andreev bound states appear within a coherence length
from the surface \cite{Hu,Tanaka95,Buchholtz,KashiwayaReport}. These
states split in the presence of a screening current
\cite{Fogelstroem,Aprili,Krupke} and they carry an anomalous
counter-flowing paramagnetic surface current \cite{Fogelstroem,Walter}. 
In previous work we have shown that the anomalous surface
current leads to a strong modification of linear
response properties \cite{Iniotakis,Zare}. Here,
we study their influence on the nonlinear Meissner effect.
We will show that the contribution of the surface Andreev
bound states to the nonlinear response coefficient follows
a $1/T^3$ law, which will ultimately dominate the bulk
$1/T$ behavior at sufficiently low temperatures. We show that the 
cross-over from bulk dominated behavior to surface dominated
behavior occurs at a
comparatively high temperature of $T/T_c \sim 1/\sqrt{\kappa}$.
This means that even for a high $\kappa \sim 100$ as is
realized in the cuprates the effect will become dominant at
temperatures below about $0.1 T_c$.

In order to calculate the nonlinear response coefficient we solve
Eilenberger's equations \cite{Eilenberger,Larkin} fully momentum and energy dependent
solving self-consistently the gap equation and the equation for
the current density 
\begin{equation}
\label{EQjQC}
\mathbf{j}(\mathbf r)=4\pi e N_0 k_B T \sum^{\omega_c}_{\varepsilon_n > 0} 
\left\langle \mathbf{v}_F(\hat{\mathbf k}) g(\mathbf r,\hat{\mathbf k},\varepsilon_n) \right\rangle_{FS}
\end{equation}
together with the Maxwell equation
\begin{equation}
\label{EQrotrot}
\nabla\times\nabla\times \mathbf{A}(\mathbf r)=\mu_0 \mathbf{j}(\mathbf r).
\end{equation}
Here, $\mathbf{A}$ is the vector potential, $\mathbf{v}_F$ is the Fermi velocity, $N_0$ the
single spin density of states, and 
$g(\mathbf r,\hat{\mathbf k},\varepsilon_n)$ the Eilenberger propagator on Matsubara
frequencies $\varepsilon_n$. The full set of equations and a description of the numerical
solution procedure based on the Riccati technique \cite{Schopohl} can be found in Ref. \onlinecite{Zare}.

We consider a homogeneous superconducting half-space in the region $x\geq 0$ with an external magnetic
field $\mathbf{B}_0$ parallel to the $z$-axis, which shall be aligned with the $c$-axis of the
crystal structure. In this geometry the current flows along the
$y$-direction in the superconductor. The gap function is assumed
to have a rotated $d$-wave form $\Delta(x,\theta) = \Delta_0(x) \cos 2\left( \theta - \alpha \right)$,
where $\alpha$ is the angle of rotation with respect to the surface and the angle $\theta$
denotes the direction of momentum within the $ab$-plane. As this problem is
translationally invariant in $y$- and $z$-direction, all quantities only depend on the
spatial variable $x$. The self-consistent solution of Eilenberger's equations on real
frequencies allows us to calculate the local, angular resolved normalized density of states
\begin{equation}
N\left(E,x,\theta \right)=-{\mathrm{Im}} \, g\left(x,\theta,i\epsilon_{n}\rightarrow E+i0^{+}\right)
\end{equation}
The equation for the $y$-component of the current density (\ref{EQjQC}) can be transformed by contour integration
and analytic continuation to the real axis \cite{Xu,Dahmimp}:
\begin{eqnarray}
j\left(x\right)&=&\frac{2}{\pi}eN_{0}v_{F}\int_{-\infty}^{\infty}dE \int_0^\pi d\theta \, \sin \theta \cdot \nonumber \\
&& f\left(E\right) \left[N_{+}\left(E,x,\theta\right)-N_{-}\left(E,x,\theta\right)\right] 
\label{jrealaxis}
\end{eqnarray}
where  $f\left(E\right)=\frac{1}{1+e^{E/T}}$
is the Fermi function and $N_{\pm}$ denotes the normalized density of states
for comoving and countermoving quasiparticles relative to the condensate
flow, i.e.
$ N_{+}\left(E,x,\theta\right) = N_{-}\left(E,x,\theta - \pi\right) $.
Once the current density distribution $j(x)$ is obtained, the magnetic field distribution
$B(x)$ and the vector potential $A(x)$ are found from integration of Eq. (\ref{EQrotrot}).

For a high-$\kappa$ superconductor the length scale of variation of the vector potential,
the magnetic penetration length $\lambda$,
is a factor of $\kappa$ larger than the variation of the Eilenberger propagator $g$ on
the length scale of the coherence length $\xi_0= \hbar v_{F}/ \pi \Delta_0$.
Thus, for temperatures $T/T_c \gtrsim 1/\kappa$ it is a very good approximation to evaluate the angular resolved local density of
states by a local Doppler shift of the fully nonlocal Eilenberger propagator in
the absence of a vector potential, i.e.
\begin{equation}
N_{\pm}\left(E,x,\theta\right) = N\left(E \pm e \mathbf{v}_F \cdot \mathbf{A}(x) ,x,\theta \right) 
\label{NDopplershift}
\end{equation}
where $N\left(E,x,\theta\right)$ on the right hand side is calculated with $\mathbf{A}(x)=0$ but fully includes the
surface Andreev bound states. Here, we have chosen the real gauge in which the vector potential is directly
proportional to the superfluid velocity $\mathbf{v}_s(x)=-\frac{e}{m} \mathbf{A}(x)$.

\begin{figure}[t]
\includegraphics[width=0.87 \columnwidth]{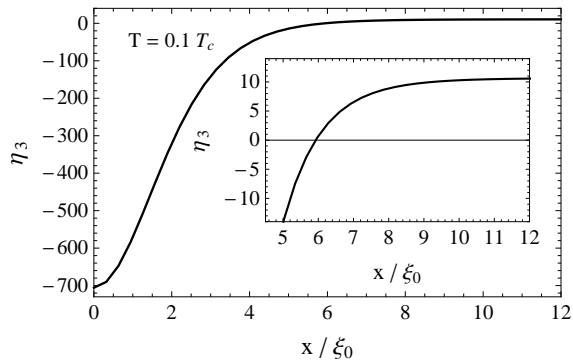}
\caption{\label{Fig01}
Spatial dependence of the nonlinear coefficient $\eta_3$ at
$T=0.1 T_c$ and $\alpha=\pi/4$ as a function of the distance $x$ from
the surface in units of the coherence length $\xi_0$. Inset: larger scale for
$x>4\xi_0$, highlighting the sign change of $\eta_3$.
}
\end{figure}

In order to determine the lowest order nonlinear response, Eq.~(\ref{NDopplershift}) is substituted into 
Eq.~(\ref{jrealaxis}) and we make a Taylor series
expansion of $j$ in the vector potential $A(x)$:
\begin{eqnarray}
j\left(x\right) &=& - 2e^2 v_F^2 N_0 \, \eta_1 \left( x \right) A(x) + \label{Taylorj}\\
&& + \frac{2e^4 v_F^4 N_0}{\Delta_0^2}
\, \eta_3  \left( x \right) A^3(x) + {\cal O}(A^5) \nonumber
\end{eqnarray}
Here, the even terms in $A$ cancel out due to symmetry. After a partial integration the 
dimensionless expansion coefficients are given by the expressions:
\begin{eqnarray}
\eta_1 &=& 1+\frac{2}{\pi}\int_{0}^{\pi}d\theta \sin^{2}\theta\int_{-\infty}^{\infty}dE
\frac{\partial f}{\partial E} N\left(E,x,\theta\right) \label{eta1} \\
\eta_3 &=& - \Delta_0^2 \frac{2}{\pi}\int_{0}^{\pi}d\theta \sin^{4}\theta\int_{-\infty}^{\infty}dE
\frac{\partial^3 f}{\partial E^3} N\left(E,x,\theta\right) 
\label{eta3}
\end{eqnarray}
where $\Delta_0$ is the zero temperature gap value in the bulk.
Note, that in contrast to the bulk calculation \cite{Dahm,Dahmimp} the expansion coefficients
now depend on the distance from the surface. Within a distance of the order of the
coherence length they contain contributions from the Andreev bound states.
The coefficient $\eta_1$ describes the linear response and the coefficient
$\eta_3$ the lowest order nonlinear response. The spatial dependence of
$\eta_3$ is shown in Fig.~\ref{Fig01} for a (110) surface ($\alpha=\pi/4$)
at a temperature of $T=0.1 T_c$. Deep in the bulk, $\eta_3$ is positive
and reaches the low temperature value $\Delta_0/2T$ known from previous work \cite{Dahm}. However,
when the surface is approached within a few coherence lengths, $\eta_3$
changes sign and reaches extremely large negative values at the surface.

\begin{figure}[t]
\includegraphics[width=0.87 \columnwidth]{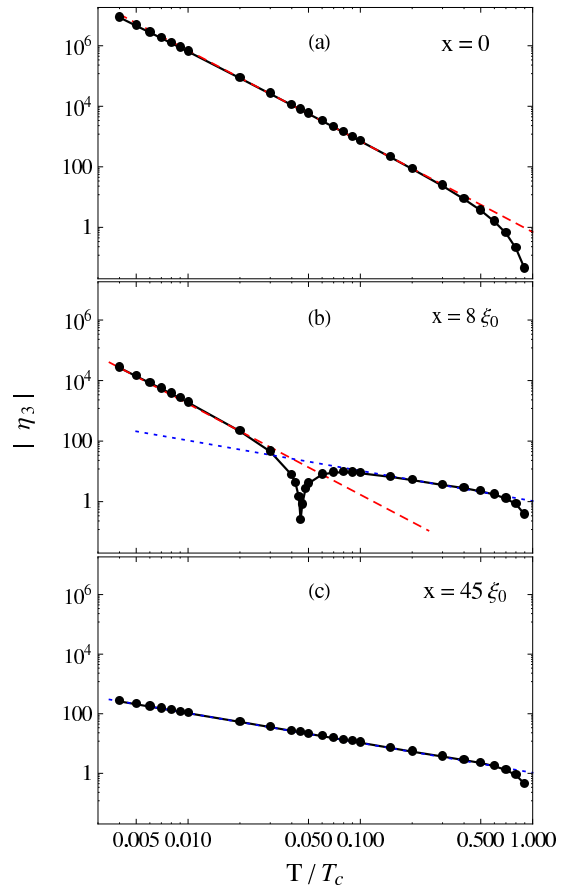}
\caption{\label{Fig02} (Color online)
Double logarithmic plot of $|\eta_3|$ as a function of temperature
$T/T_c$ for three selected positions: (a) $x=0$, (b) $x=8 \xi_0$,
and (c) $x=45 \xi_0$. The red dashed lines show a $1/T^3$ dependence
and the blue dotted lines a $1/T$ dependence.
}
\end{figure}

The temperature dependence of the modulus $|\eta_3|$ is shown in
Fig.~\ref{Fig02} on a double logarithmic scale for three selected
spacial positions. Fig.~\ref{Fig02}(c) shows the temperature
dependence at $x=45 \xi_0$ in the bulk. As is well known
from previous work, $|\eta_3|$ follows a $1/T$ law at low temperatures
(blue dotted line). Right at the surface ($x=0$), however, Fig.~\ref{Fig02}(a)
demonstrates that $|\eta_3|$ is following a $1/T^3$ behavior (red dashed line).
In Fig.~\ref{Fig02}(b) an intermediate position at $x=8 \xi_0$
is shown. In this case, at higher temperatures a $1/T$ law is followed.
At a certain temperature, $\eta_3$ changes sign and starts to follow
a $1/T^3$ behavior below that temperature.
These results clearly show that the nonlinear response coming from the
surface area, where the Andreev bound states are present, is much
stronger and of opposite sign than the nonlinear response in the bulk.

In a typical intermodulation experiment only the total response of
the system is probed. The quantity that is observed is the nonlinear
change of the total inductance of the system \cite{Dahm}. The total
inductance $L$ can be calculated from the total kinetic
and magnetic field energy in the system via the equation
\begin{equation}
\frac{1}{2} L I^2 = \frac{1}{2\mu_0} \int_0^\infty dx \left( B^2 \left(x\right) - \mu_0 j\left(x\right) A\left(x\right) \right)
\label{fieldenergy}
\end{equation}
where $I=\int_0^\infty dx \, j\left(x\right) $
is the total current per unit length \cite{Dahm}.
Using Eq.~(\ref{EQrotrot}), $B=dA/dx$, and the fact that the
magnetic field vanishes in the bulk, Eq.~(\ref{fieldenergy})
can be brought by partial integration into the more convenient form
\begin{equation}
L = - \frac{A_0}{I}
\label{inductance}
\end{equation}
with $A_0 = A(x=0)$. To lowest order in $A_0$ the total current $I$
generally will be of the form
\begin{equation}
I = a_1 A_0 + a_3 A_0^3
\label{IofA}
\end{equation}
The intermodulation response is proportional to the nonlinear coefficient
$\left. \frac{\partial^2 L}{\partial I^2} \right|_{I=0}$ \cite{Dahm}.
A straightforward calculation shows that this quantity can be related to the
expansion coefficients $a_1$ and $a_3$ using Eq.~(\ref{inductance}) 
\begin{equation}
\left. \frac{\partial^2 L}{\partial I^2} \right|_{I=0} = \frac{2 a_3}{a_1^4}
\end{equation}

\begin{figure}[t]
\includegraphics[width=0.87 \columnwidth]{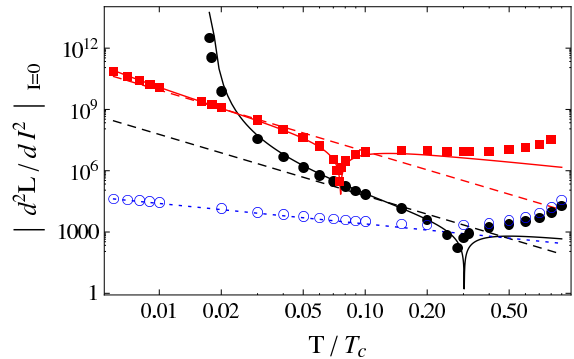}
\caption{\label{Fig03} (Color online)
Double logarithmic plot of $\left| \frac{\partial^2 L}{\partial I^2} \right|_{I=0}$ 
as a function of temperature $T/T_c$ for $\kappa=63$ and
$\alpha=\pi/4$ (solid black circles), for $\kappa=1000$ and $\alpha=\pi/4$ (solid red squares), and
for $\kappa=63$ and $\alpha=0$ (open blue circles). The dashed lines show a
$1/T^3$ behavior and the dotted line a $1/T$ behavior. The solid lines
show the approximation Eq.~(\ref{d2Lapprox}) for $\kappa=63$ (black) and $\kappa=1000$ (red),
respectively.
}
\end{figure}

We have determined $a_1$ and $a_3$ from our numerical solution of Eilenberger's
equations. The resulting values for $\left| \frac{\partial^2 L}{\partial I^2} \right|_{I=0}$
are shown in Fig.~\ref{Fig03} as a function of reduced temperature for $\kappa=63$
(solid black circles) and $\kappa=1000$ (solid red squares) on a double logarithmic scale. 
Decreasing the temperature from $T_c$, for $\alpha=\pi/4$
the nonlinear coefficient initially decreases and changes sign at a temperature
near $T/T_c \approx 2.4/\sqrt{\kappa}$. Below that temperature the nonlinear
coefficient increases following a $1/T^3$ law and finally diverges at a temperature
near $T/T_c \approx 1/\kappa$. For comparison also the behavior for $\kappa=63$ and
$\alpha=0$ is shown, when the surface states are absent (open circles). In this case there is
no sign change and the nonlinear coefficient follows a $1/T$ behavior at low
temperatures, as known from the bulk.

In order to check the validity of the numerical calculations and obtain a physical
understanding of the results, we made an approximate analytical solution of the
problem which we present now. For a piecewise constant gap function Eilenberger's
equations can be solved analytically \cite{Schopohl}. As an approximation we
assume that the $d$-wave gap is constant in space. Then, the analytical solution
of Eilenberger's equations allows us to determine the residue of the zero energy
pole of the Eilenberger propagator analytically, which contains the contributions
from the zero energy bound states at the surface. As a result, we find the
following expression for the bound state contribution to the local, angular
resolved density of states for $\alpha=\pi/4$ in the absence of an external
field:
\begin{equation}
N_{bs}\left(E,x,\theta\right) = \pi \Delta_0 \left| \sin 2\theta \right|
e^{- \frac{4}{\pi}\left| \sin \theta \right| \frac{x}{\xi_0} } \delta\left( E \right)
\label{Nboundstate}
\end{equation}
The $\delta$-function shows that the bound states are only present at
zero energy. The exponential factor drops off on a length scale of the
coherence length, showing that these states are localized at
the surface. Introducing this expression into Eq.~(\ref{eta3})
the energy integration immediately shows that the bound states
lead to a $1/T^3$ scaling, which is of
opposite sign than the bulk behavior, because $\frac{\partial^3 f}{\partial E^3}$
is positive at zero energy, but negative at higher energies.

In order to determine the coefficients $a_1$ and $a_3$ in Eq.~(\ref{IofA}), we
integrate Eq.~(\ref{Taylorj}) using the following approximations. For $\kappa = \lambda/\xi_0 \gg 1$ 
we can assume that the vector potential varies exponentially on the length scale
of the penetration length $\lambda$, and make the ansatz 
\begin{equation}
A(x)=(A_0-\epsilon) e^{-x/\lambda}+\epsilon \, e^{-3 x/\lambda}
\label{Aansatz}
\end{equation} 
The functions
$\eta_1$ and $\eta_3$ both vary on the length scale of the coherence length, which
is much smaller than $\lambda$. Therefore, we can approximate them as
\begin{eqnarray*}
\eta_1 (x) &=& c_1 \delta (x) + \eta_{1b} \\
\eta_3 (x) &=& c_3 \delta (x) + \eta_{3b} 
\end{eqnarray*}
Here, $\eta_{1b}$ and $\eta_{3b}$ are the bulk values of $\eta_1$ and $\eta_3$, respectively.
The coefficients $c_1$ and $c_3$ describe the contributions of the surface bound states.
They are obtained by substituting Eq.~(\ref{Nboundstate}) into Eq.~(\ref{eta1}) 
and Eq.~(\ref{eta3}) and integrating over $x$ from 0 to $\infty$. This yields
$ c_1 =  - \frac{\pi \Delta_0}{6 T} \xi_0$ and $c_3 =  - \frac{\pi \Delta_0^3}{20 T^3} \xi_0$.
The parameter $\epsilon$ in Eq.~(\ref{Aansatz}) is determined from the differential equation
Eq.~(\ref{EQrotrot}) together with Eq.~(\ref{Taylorj}) and up to order
$A_0^3$ found to be
\begin{equation}
\epsilon = \frac{1}{8} \frac{e^2v_F^2 \eta_{3b}}{\Delta_0^2 \eta_{1b}} A_0^3 . 
\end{equation}
With these approximations we find from integrated Eq.~(\ref{Taylorj})
\begin{eqnarray*}
a_1 &=&  - 2e^2 v_F^2 N_0 \left( c_1 + \lambda \eta_{1b} \right) \\
a_3 &=&  \frac{2e^4 v_F^4 N_0}{\Delta_0^2} \left( c_3 + \frac{\lambda}{4} \eta_{3b} \right)
\end{eqnarray*}
Using the low temperature limiting expressions $\eta_{1b} \sim 1$ and
$\eta_{3b} \sim \Delta_0/2T$ finally leads to
\begin{equation}
\left. \frac{\partial^2 L}{\partial I^2} \right|_{I=0} = 
\frac{1}{16 e^4 v_F^4 N_0^3 \Delta_0^2 \lambda^3} \frac{\frac{1}{2} \frac{\Delta_0}{T}-
\frac{\pi}{5}\frac{1}{\kappa}\frac{\Delta_0^3}{T^3}}{\left( 1 - \frac{\pi}{6} \frac{1}{\kappa} 
\frac{\Delta_0}{T} \right)^4}
\label{d2Lapprox}
\end{equation}
This expression shows that upon lowering the temperature
from $T_c$ the total nonlinear response of the system
initially follows the $1/T$ increase caused by the
bulk nonlinearities (first term in the numerator). 
At a temperature of the order of
$T/T_c \sim 1/\sqrt{\kappa}$ the nonlinearities of
the surface states become comparable with the bulk
contributions and cancel them (second term in the numerator). 
Below that temperature
the $1/T^3$ increase with opposite sign dominates
due to the surface states. Finally, at a temperature
of the order of $T/T_c \sim 1/\kappa$ the nonlinear
response diverges (denominator). This divergence signals the
breakdown of the large $\kappa$ approximation we have used 
here. The approximate expression Eq.~(\ref{d2Lapprox})
is shown in Fig.~\ref{Fig03} together with the numerical results.
The agreement is quite good at low temperatures despite the approximations made.

To conclude, we have shown that in a $d$-wave superconductor
surface Andreev bound states lead to a strong contribution
to the nonlinear Meissner effect, which follows a $1/T^3$
behavior at low temperatures and is of opposite sign
compared to the bulk nonlinear response. At temperatures
below $T/T_c \sim 1/\sqrt{\kappa}$ these contributions
dominate the total nonlinear response. Such temperatures
are readily available in intermodulation experiments and make
them a tool to study surface Andreev bound states.
The fingerprint of the Andreev bound states should be a
$1/T^3$ temperature dependence and a sign change
($180^\circ$ relative phase change) in the nonlinear part
of the inductance. So far, intermodulation experiments
have been mostly done on systems with (100) surfaces,
where Andreev bound states are absent. In systems with
(110) surfaces the effect studied here should become
most prominent.

This work was supported by the Deutsche Forschungsgemeinschaft under
grant No. Da~514/2.

\end{document}